\documentclass[aip,jcp,reprint,amsmath,amssymb]{revtex4-2}
\usepackage{graphicx}
\usepackage[suffix=]{epstopdf}
\usepackage{natmove}
\begin{document}
\title{Multiscale modeling of solute diffusion in triblock copolymer membranes}
\author{Anthony J. Cooper}
\affiliation{Department of Physics, University of California, Santa Barbara, California 93106, USA}

\author{Michael P. Howard}
\affiliation{McKetta Department of Chemical Engineering, University of Texas at Austin, Austin, Texas 78712, USA}
\affiliation{Department of Chemical Engineering, Auburn University, Auburn, Alabama 36849, USA}

\author{Sanket Kadulkar}
\affiliation{McKetta Department of Chemical Engineering, University of Texas at Austin, Austin, Texas 78712, USA}

\author{David Zhao}
\affiliation{McKetta Department of Chemical Engineering, University of Texas at Austin, Austin, Texas 78712, USA}
\affiliation{Department of Chemical Engineering, University of California, Santa Barbara, California 93106, USA}

\author{Kris T. Delaney}
\affiliation{Materials Research Laboratory, University of California, Santa Barbara, California 93106, USA}

\author{Venkat Ganesan}
\affiliation{McKetta Department of Chemical Engineering, University of Texas at Austin, Austin, Texas 78712, USA}

\author{Thomas M. Truskett}
\email{truskett@che.utexas.edu}
\affiliation{McKetta Department of Chemical Engineering, University of Texas at Austin, Austin, Texas 78712, USA}
\affiliation{Department of Physics, University of Texas at Austin, Austin, Texas 78712, USA}

\author{Glenn H. Fredrickson}
\email{ghf@ucsb.edu}
\affiliation{Materials Research Laboratory, University of California, Santa Barbara, California 93106, USA}
\affiliation{Department of Chemical Engineering, University of California, Santa Barbara, California 93106, USA}
\affiliation{Materials Department, University of California, Santa Barbara, California 93106, USA}

\begin{abstract}
We develop a multiscale simulation model for diffusion of solutes through porous triblock copolymer membranes. The approach combines two techniques: self-consistent field theory (SCFT) to predict the structure of the self-assembled, solvated membrane and on-lattice kinetic Monte Carlo (kMC) simulations to model diffusion of solutes. Solvation is simulated in SCFT by constraining the glassy membrane matrix while relaxing the brush-like membrane pore coating against the solvent. The kMC simulations capture the resulting solute spatial distribution and concentration-dependent local diffusivity in the polymer-coated pores; we parameterize the latter using particle-based simulations. We apply our approach to simulate solute diffusion through nonequilibrium morphologies of a model triblock copolymer, and we correlate diffusivity with structural descriptors of the morphologies. We also compare the model's predictions to alternative approaches based on simple lattice random walks and find our multiscale model to be more robust and systematic to parameterize. Our multiscale modeling approach is general and can be readily extended in the future to other chemistries, morphologies, and models for the local solute diffusivity and interactions with the membrane.
\end{abstract}

\maketitle

\section{Introduction}
Block copolymers self-assemble into microphase-separated ordered structures \cite{Bates1990, BATES898}, such as hexagonally packed cylinders, that can produce isoporous membranes with higher selectivity and permeability compared to membranes made from homopolymers \cite{Zhang2018, Abetz2015}. The pore diameter is important for engineering membranes that employ a sieving mechanism for filtration. The membrane pore diameters can be controlled through the composition and molecular weight of the polymer, by incorporating additives, or by using a block copolymer blend \cite{Gu2015, Radjabian2015}. The pores can also be chemically functionalized, such as with solute-selective ligands, to further improve separation performance \cite{Sujanani2020}. As a result, block copolymers are promising materials for fabricating membranes, including mesoscopic membranes used for gas separation \cite{gassep_BC} and ultrafiltration membranes used for water filtration \cite{waterfilt_BC, Zhang2018}.

There are a number of ways to fabricate block copolymer membranes.  One method is to spincoat a block copolymer solution onto a substrate and anneal it, allowing the polymers to relax toward their equilibrium state \cite{Sinturel2013}. The final structure can be predicted with knowledge of only the enthalpic interactions (i.e., Flory--Huggins parameters) and the molecular weights of the constituent blocks \cite{Moon2020}. Pores are then created by etching a sacrificial block, while another block acts as the membrane matrix.  Another prominent method combines the highly scalable process of nonsolvent induced phase separation, typically used to form homopolymer membranes, and the self-assembly of block copolymers initiated by the evaporation of the solvent (SNIPS) \cite{Peinemann2007, Nunes2016, Wang2012, Woo2017}. Although more scalable, controlling the membrane morphology in a SNIPS process poses a significant challenge because nonequilibrium structures can form, including transient  percolation networks or spinodal networks \cite{Stegelmeier2014}.  

Diblock copolymers have often been used to make membranes because they are relatively simple to synthesize and have a well-known phase diagram \cite{Cochran2006}. Commonly, one block is chosen to be a glassy material, such as polystyrene, that forms a solid matrix, while the other is a sacrificial material, such as poly(lactic acid), that can be etched to form pores. However, triblock copolymers have recently garnered attention \cite{Rzayev2005, PhillipW2011, Dorin2012, Xu2020, Zhang2017} because the additional architectural and chemical complexity of the polymer offers greater tunability of the membrane. For example, a hydrophilic block, such as poly(ethylene oxide), can be inserted between the glassy matrix block and sacrificial block in order to coat the pores and improve water uptake; additional chemical moities can also be added to this block to improve selectivity \cite{Zhang2017}.  Furthermore, the inclusion of another block in the matrix, e.g., attaching polyisoprene onto poly(styrene-b-4-vinylpyridine) to make poly(isoprene-b-styrene-b-4-vinylpyridine), has been shown to improve mechanical stability \cite{PhillipW2011, Xu2020}. Triblock copolymers also have an expanded phase space that can lead to advantageous morphologies not seen in diblocks \cite{PhillipW2011} and widen the phase window of bicontinuous morphologies with improved toughness \cite{Dair1999}. These bicontinuous structures also do not require alignment of the domains, unlike cylindrical morphologies. However, as the number of possible morphologies expands, it becomes increasingly necessary to determine which are optimal (or even suitable) for filtration and to understand the relationship between morphology and diffusive transport of solutes through the pores.

Fickian diffusion through cylindrical and lamellar pores is well known to be the one- and two-dimensional equivalent of bulk diffusion, respectively. However, diffusion through real porous media can be complicated by issues such as molecular interactions and confinement \cite{Tartakovsky2019}. Experimentally characterizing diffusion through membrane structures is challenging due to the need to fabricate the structures and measure transport over the required length scales \cite{Zalarmi2018}.  As a result, a variety of simulation studies have been performed to investigate an analogous problem of ion self-diffusion through conducting block copolymers \cite{Shen2018, Alshammasi2018, Venkat2021}. Shen \textit{et al}.\cite{Shen2018} and Alshammasi and Escobedo\cite{Alshammasi2018} both simulated the transport of ions through common block copolymer morphologies (lamellar, cylindrical and gyroid) and found that diffusion through the gyroid network is slower than through oriented lamellae because the increased tortuosity of the network hampers diffusion more than three-dimensional continuity and percolation promotes it. On the other hand, Zhang \textit{et al}.\cite{Venkat2021} found that once dimensionality was taken into account, the morphology had little influence over the anion diffusivity through block copolymeric ionic liquids and instead depended on the concentration of interfacial anions.

Although these studies provide great insight into the structure--transport relationship for various block copolymer morphologies, they have all relied on either simple random-walk diffusion models that miss do not capture certain pore-scale effects or more detailed particle-based simulations that cannot easily access the length scales of some membrane morphologies. For example, Howard \textit{et al}. recently combined self-consistent field theory (SCFT), which was used to generate block copolymer membrane morphologies, with a lattice-based random-walk transport model to study the self-diffusion of a solute \cite{Howard2020}. To further characterize the important features of each morphology, structural descriptors were calculated and random-forest regression was used to investigate which descriptors had greatest correlation with the self-diffusivity. However, this study neglected the effect of the polymers coating the pores and focused solely on the morphology of the glassy matrix. Pore coatings have been shown to play an important role in water transport through lamellar and cylindrical pores using particle-based dissipative particle dynamics (DPD) simulations \cite{Dipak2020}; however, these simulation techniques are too computationally demanding to apply to more complex membrane morphologies. To this end, it would be highly beneficial to have an efficient high-throughput method for faithfully modeling the diffusion of a solute through various block copolymer structures.

In this study, we develop a multiscale simulation framework, built on our prior work \cite{Howard2020}, that incorporates the effect of the pore coating on the self-diffusion of solutes through triblock copolymer membranes.  Specifically, we focus on nonequilibrium membrane morphologies that are difficult to simulate using more detailed models like DPD.  In order to do this, we present a novel SCFT method for simulating the structural effect of the solvent on the pore coating, then use an on-lattice kinetic Monte Carlo (kMC) model accounting for obstruction from the pore coating to simulate solute diffusion through the pores of the SCFT-generated morphologies. The kMC model uses local pore-level diffusion data from more detailed DPD simulations to model membrane-scale diffusion at much larger length and time scales than are accessible in DPD.  Our use of kMC is partially motivated by its recent success in capturing the impact of shale rock features on fluid transport where discrete regimes were used to distinguish between the center of the porous region and the interface \cite{Apostolopoulo2019, Apostolopoulou2021}.  After determining the self-diffusion coefficient from kMC simulations, we compare the results of our more detailed model with our prior work using simple random-walk simulations based on two definitions of the pore.  We show that including the pore coating leads to qualitatively different trends in diffusivity as a function of block fractions, which we correlate with various structural descriptors of the morphology. We find that the features that are important for predicting the solute diffusivity are vastly different when effects of the pore coating are taken into account.

\begin{figure}
    \centering
    \includegraphics{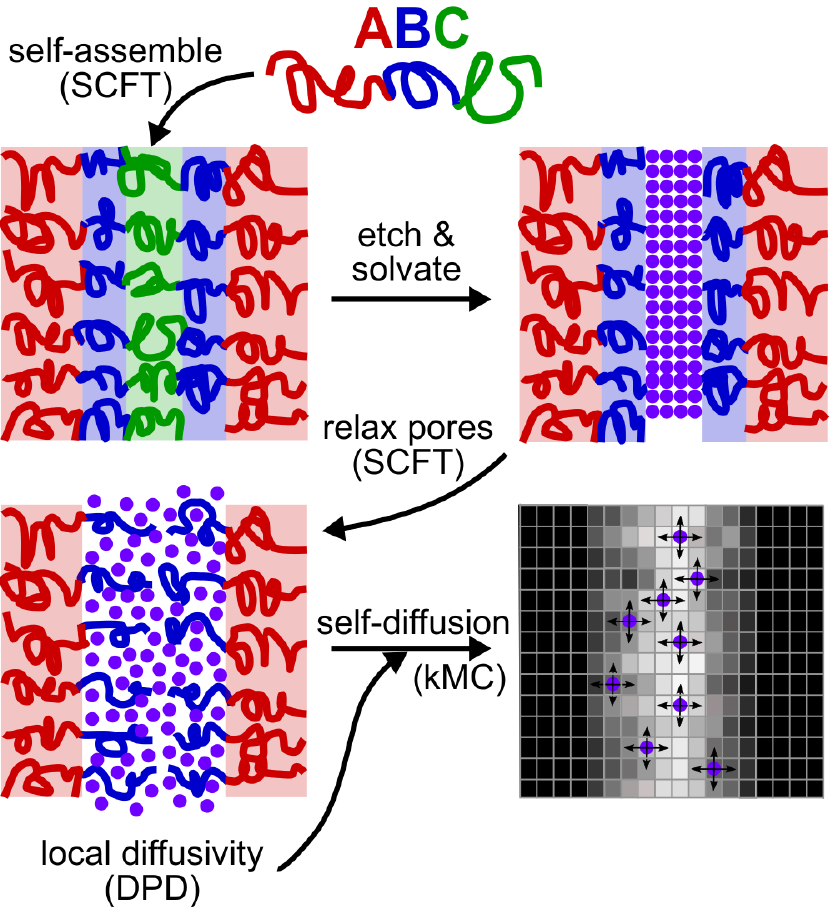}
    \caption{Multiscale simulation workflow used to study solute diffusion through the solvated pores of the self-assembled ABC triblock copolymer membranes. First, SCFT is used to model self-assembly of the ABC triblock copolymer. The C block is then etched out and replaced with solvent. Another SCFT simulation is  run to relax the pore-coating B block and solvent while constraining the glassy A block to maintain the initial matrix morphology. The resulting density distributions are input to the kMC model, along with a model for the local diffusivity computed with DPD simulations, to obtain the diffusivity of solute tracers through the pores.}
    \label{fig:workflow}
\end{figure}

\section{Computational framework}
\label{sec:method}
We studied membrane morphologies made from a model ABC triblock copolymer (Fig.~\ref{fig:workflow}) having $\chi_{\text{AC}}N = 35$ and $\chi_{\text{AB}}N = \chi_{\text{BC}}N = 13$, where $\chi_{ij}$ is the Flory--Huggins interaction parameter between blocks $i$ and $j$, and $N$ is the overall degree of polymerization. Tyler \textit{et al}.~computed the equilibrium phase diagram of this idealized nonfrustrated triblock copolymer as a function of the overall block volume fractions $f_{\rm A}$, $f_{\rm B}$, and $f_{\rm C}$ using SCFT \cite{Tyler2007}. The majority of phases discovered were core--shell analogues of the structures found in diblocks such as lamellae and hexagonally packed cylinders, with one end block (A or C) acting as the core, the middle block (B) acting as the shell, and the other end block (C or A) forming a continuous domain. These structures are of particular interest in membrane fabrication because the core-forming block can be sacrificed, e.g., using etching, to form pores coated by the middle block embedded within a self-supporting continuous matrix. For this study, we designated the A block as a glassy block that forms the membrane matrix and the C block as the sacrificial block to be removed.

In order to study the diffusion of a solute through the porous membrane, we developed and applied a multiscale modeling workflow (Fig.~\ref{fig:workflow}). First, the self-assembly of the triblock copolymer is simulated using SCFT. We then ``etch'' out the C block and replace it with solvent. In order to obtain the distribution of the solvent and pore-coating B block in the newly formed pores, we run an additional SCFT simulation to reach a locally stable structure while constraining the density profile of the glassy A block so that the matrix morphology does not change. Finally, the solute density distributions from SCFT and the solute dynamics computed from pore-level DPD simulations are input into a kMC simulation. The solute trajectories resulting from the kMC simulations are used to obtain the self-diffusivity of the solute through the porous network. We describe each of these steps in detail.

\subsection{Membrane morphology}
\label{sec:method:scft}
Six self-assembled morphologies per triblock composition were selected from the nonequilibrium morphologies generated by Howard \textit{et al}. \cite{Howard2020} to facilitate direct comparisons between that study and this one.  Details on how these morphologies were generated can be found in their publication \cite{Howard2020}. We focus here only on morphologies with cubic simulation cells of length $16 R_{\rm g}$, where $R_{\rm g}=b(N/6)^{1/2}$ is the radius of gyration of an ideal polymer chain and $b$ is the statistical segment length of the triblock copolymer. The simulation cell had periodic boundary conditions in all three dimensions. From the melt morphologies, we removed the C block and replaced it with a solvent (S) that we modeled in SCFT as a ``point'' polymer, while also choosing $N = 100$ as the nominal degree of polymerization of the triblock copolymer. After etching the C block, the remaining AB diblock had degree of polymerization $N^{\rm AB} = (f_{\rm A} + f_{\rm B}) N$. The relative fraction of block $i\in \{A,B\}$ within the polymer was $f_i/(f_{\rm A} + f_{\rm B})$, but the overall block volume fractions remained the same. The Flory--Huggins interaction parameters between the polymer and solvent were chosen to be $\chi_{\rm AS} = 2$ and $\chi_{\rm BS} = 0$ in order to model a hydrophobic A block and a hydrophilic B block (assuming S is water-like). 

To mimic the experimental process of solvating the newly formed membrane pores, SCFT simulations were then performed to relax the B block and solvent to a local free-energy minimum while effectively ``freezing'' the glassy A block so that the overall membrane morphology did not change. In order to freeze the A block, a novel method was developed and implemented into our in-house SCFT software. A harmonic energy penalty $U$ for the A species was added to the Hamiltonian,
\begin{equation}
    \beta U = \frac{\kappa}{2\rho_0} \int {\rm d}\mathbf{r} \ (\rho_{\rm A}(\mathbf{r}) - \rho_{{\rm A},{\rm t}}(\mathbf{r}))^2,
\end{equation}
where $\rho_{\rm A}$ is the density profile of the A block, $\rho_{{\rm A},{\rm t}}$ is the target density profile of the A block (i.e., the output from the original SCFT simulation), and $\kappa$ determines the strength of the penalty. Moreover, $\beta = 1/(k_{\rm B} T)$, $k_{\rm B}$ is the Boltzmann constant, $T$ is the temperature, and $\rho_0$ is the total monomer density.  For sufficiently large values of $\kappa$, the ultimate A density profile becomes quantitatively indistinguishable from that obtained in the original SCFT simulation.

The diblock and solvent (AB+S) system was represented using the incompressible multispecies exchange model \cite{Duchs2014} with Gaussian chain statistics, and the exchange-mapped chemical potential fields for the AB+S system were initialized using those obtained from the ABC morphologies. The semi-implicit Siedel scheme \cite{Ceniceros2004} was used to perform field updates and the modified diffusion equation was solved using a second-order operator-splitting algorithm \cite{Rasmussen2002} with contour stepping $\Delta s = 0.01N$. A simulation with $\kappa = 600$ was run until numerical convergence, then the result was used to initialize simulations with larger $\kappa$. This process of chaining together simulations with increasing $\kappa$ was continued until the root-mean-square difference between the actual and target density profiles was no greater than $0.01\,R_{\rm g}^{-3}$ at any location in the cell.

\subsection{Membrane-scale diffusion}
\label{sec:method:kmc}
Once the morphologies were generated, we employed a continuum approach to model diffusive motion through the membrane. We assumed that the solute undergoes Brownian motion that depends on its position. For example, diffusion can be slower in regions of higher polymer concentration where motion is more obstructed. For simplicity, we further assumed that the rate of diffusion was similar in all directions so that solute motion was characterized by a scalar diffusion coefficient $D(\mathbf{r})$ at each position $\mathbf{r}$; this assumption is reasonable for small solutes that are not too close to a surface. To account for influence of the membrane morphology on the solute distribution, we further considered an effective external field $\psi(\mathbf{r})$ that acted on the solutes.

Under these assumptions, the probability density $\rho(\mathbf{r},t)$ to find a tracer solute at position $\mathbf{r}$ and time $t$ then follows a conservation law (Fokker--Planck equation) \cite{Howard2016},
\begin{equation}
\frac{\partial \rho(\mathbf{r},t)}{\partial t} = \nabla \cdot\left[D(\mathbf{r}) \rho(\mathbf{r}) \beta \nabla \psi(\mathbf{r}) + D(\mathbf{r})\nabla\rho(\mathbf{r}) \right].
\label{eq:fp}
\end{equation}
The initial condition $\rho(\mathbf{r},0) = \delta(\mathbf{r}-\mathbf{r}_0)$ was based on the initial position $\mathbf{r}_0$ of the tracer at $t=0$, while the boundary conditions were periodic for the morphology, as in the SCFT calculations. At long times, $\rho$ evolves toward a steady state $\rho_\infty(\mathbf{r}) \sim \exp[-\beta\psi(\mathbf{r})]$, so $\psi$ can be chosen to achieve a targeted equilibrium distribution. Note that Eq.~\eqref{eq:fp} can equivalently be written as
\begin{equation}
\frac{\partial \rho}{\partial t} = \nabla\cdot\left[\rho_\infty D \nabla\left(\frac{\rho}{\rho_\infty}\right)\right],
\label{eq:fpsteady}
\end{equation}
where we have omitted the arguments $\mathbf{r}$ and $t$ for notational convenience. To solve Eq.~\eqref{eq:fp}, we simulated the motion of a tracer on a lattice using a kinetic Monte Carlo (kMC) method.\cite{Magna1999, VonToussaint2015, Kadulkar2019, doi:10.1021/acs.jpclett.0c01937} The membrane morphology was discretized into a cubic lattice with edge length $\ell$, and a tracer was assumed to occupy one lattice site. A tracer starting on site $i$ was allowed to hop along Cartesian axis $\alpha$ to an adjacent site $i_\alpha^\pm$ in either the forward ($+$) or reverse ($-$) direction with a rate $k(i_\alpha^\pm |i)$. The evolution of the probability density $\rho(i,t)$ to find a tracer at site $i$ at time $t$ [the discrete equivalent of $\rho(\mathbf{r},t)$] is characterized by a master equation for this stochastic process \cite{RuizBarlett2011,RuizBarlett2013},
\begin{align}
\frac{\partial \rho(i,t)}{\partial t} = &\sum_\alpha k(i|i_\alpha^-) \rho(i_\alpha^-,t) + k(i|i_\alpha^+) \rho(i_\alpha^+,t) \nonumber\\
&- [k(i_\alpha^-|i) + k(i_\alpha^+|i)]\rho(i,t).
\label{eq:master}
\end{align}

To choose the move rates, we first imposed detailed balance $k(i_\alpha^\pm|i)\rho_\infty(i) = k(i|i_\alpha^\pm)\rho_\infty(i_\alpha^\pm)$ using the steady-state distribution $\rho_\infty$. It can be shown that in the limit of small $\ell$, Eq.~\eqref{eq:master} then approximates
\begin{equation}
\frac{\partial \rho}{\partial t} = \rho_\infty \sum_\alpha D_\alpha \frac{\partial^2}{\partial r_\alpha^2} \left(\frac{\rho}{\rho_\infty}\right) + v_\alpha \frac{\partial}{\partial r_\alpha} \left(\frac{\rho}{\rho_\infty}\right),
\label{eq:fpkmc}
\end{equation}
where the derivatives are taken with respect to the $\alpha$-component of the position coordinate, $r_\alpha$. The diffusivity $D_\alpha$ along direction $\alpha$ is defined in terms of the hopping rates at site $i$ by
\begin{equation}
D_\alpha(i) = \frac{\ell^2}{2}[k(i_\alpha^+|i)+k(i_\alpha^-|i)],
\end{equation}
while $v_\alpha$ is an effective advection along $\alpha$ defined in terms of the hopping rates at site $i$ by
\begin{equation}
v_\alpha(i) = \ell [k(i_\alpha^+|i)-k(i_\alpha^-|i)].
\end{equation}
Note that because $D$ was assumed to be a scalar, $D_\alpha$ and the sum of the hopping rates at site $i$ must be equal in all directions but $v_\alpha$ need not be equal. We compared Eq.~\eqref{eq:fpkmc} to Eq.~\eqref{eq:fpsteady} and chose $v_\alpha$ to make the two equivalent. This determined the hopping rates as
\begin{equation}
k(i_\alpha^\pm|i) = \frac{D(i)}{\ell^2} \pm \frac{1}{2\ell}\left(\frac{\partial D(i)}{\partial r_\alpha} - D(i) \beta \frac{\partial \psi(i)}{\partial r_\alpha} \right),
\label{eq:kmcrate}
\end{equation}
where the partial derivatives are evaluated at $i$.

Our derivation is general to any external field $\psi(\mathbf{r})$ or scalar diffusivity $D(\mathbf{r})$. The external field enforces the internal membrane morphology (e.g., regions excluded to the solute) that can be determined from measurements or simulations. It can also incorporate interactions with the membrane, such as effective attraction due to chemical functionalization. The diffusivity can be estimated from experiments, empirical diffusion models, or more detailed computer simulations. In the next section, we will describe the model solute dynamics that we studied in this work; however, we emphasize that the framework can be readily extended to incorporate other data sources.

\subsection{Pore-scale diffusion}
\label{sec:method:dpd}
We applied the kMC approach to study the diffusion of a solute tracer through the various membrane morphologies prepared in Sec.~\ref{sec:method:scft}. For convenience, we assumed that the solute tracer was chemically similar to the solvent, so its steady-state distribution $\rho_\infty$ was directly proportional to $\phi_{\rm S}(\mathbf{r})$ as determined by SCFT, and we defined $\beta \psi(\mathbf{r}) = -\ln\phi_{\rm S}(\mathbf{r})$. Note that this amounts to a potential that excludes the solute tracer from the walls of the pores. To determine $D(\mathbf{r})$, we performed DPD simulations \cite{Hoogerbrugge1992,Espanol1995,Groot:1997} of diffusion through a single lamellar pore. DPD is a mesoscopic particle-based simulation technique that has been widely used to study block copolymers. Some of us recently used DPD to study pore-level diffusion of water in triblock copolymer membranes \cite{Dipak2020}, revealing that interactions between water and the polymers inside the pore lead to slower local diffusion (i.e., at short times) in regions of higher polymer concentration. This decrease in the local diffusivity leads to a commensurate decrease in the average water diffusion at long times. Here, we used DPD simulations to measure the local tracer diffusivity $D(\mathbf{r})$ that we input to the kMC model, but other molecular modeling approaches could also be applied.

We first constructed a DPD model for the ABC triblock copolymers studied using SCFT. The polymers were represented as linear chains of $N = 100$ beads with mass $m$ and nominal diameter $d$ connected by springs; each bead was assigned a type (A, B, or C) according to its block. All beads interacted through the standard DPD conservative, random, and dissipative forces \cite{Groot:1997}. The DPD repulsion parameter between beads with the same type was $a_{ii} = 75\,k_{\rm B}T/d$, while the DPD repulsion parameter between beads with different types $a_{ij}$ was chosen to achieve the desired $\chi_{ij}N$ for the model (see below). The DPD friction parameter for all beads was $\gamma_i = 4.5\,m/\tau$, where $\tau = \sqrt{\beta m d^2}$ is the unit of time. In addition to the standard DPD forces, bonded beads additionally interacted through a harmonic potential $u_{\rm b}(r) = k(r-r_0)^2/2$ with spring constant $k = 100\,k_{\rm B}T/d^2$ and $r_0 = 1.0\,d$. All simulations were performed using HOOMD-blue (version 2.6.0) with features extended using azplugins (version 0.8.0) \cite{Anderson2020,Phillips2011,azplugins}. The integration time step was $0.01\,\tau$, and the bead number density was $3/d^3$.

In order to map length scales between the DPD model and SCFT calculations, we first prepared a homopolymer melt and measured the radius of gyration $R_{\rm g}$ and end-to-end distance $R_{\rm e}$, finding $\langle R_{\rm g}^2 \rangle^{1/2} = 4.466\,d$ and $\langle R_{\rm e}^2 \rangle^{1/2} = 10.92\,d$. Both measurements are consistent with an ideal chain having an effective segment length $b = 1.09\,d$ \cite{Rubinstein}, which is slightly larger than the nominal bead diameter. We then carried out the analysis outlined by Groot and Warren to connect $a_{ij}$ approximately to $\chi_{ij} N$ for our bead--spring model \cite{Groot:1997}. We performed direct coexistence simulations of two homopolymer oligomers having different bead types, varying the length of the oligomers from 2 to 6 beads and the difference in the repulsion parameter for unlike and like beads $\Delta a_{ij} = a_{ij}-a_{ii}$ in the range $10 \le \Delta a_{ij} N \le 30$. We initialized equal-sized slabs of each oligomer in an orthorhombic box with square cross section (edge length $10\,d$) and length $30\,d$ then allowed the mixture to equilibrate for $10^5\,\tau$, during which time the initially separated oligomers partially dissolved in each other. We then sampled configurations every $100\,\tau$ during a $10^5\,\tau$ production simulation. We computed the average bead density profile with center-of-mass shifting using a bin spacing of $0.5\,d$ \cite{Silmore2017}, extracted the coexistence volume fractions from the bulk region of each slab, and used Flory--Huggins theory to determine $\chi_{ij} N$ \cite{Rubinstein}. As in Groot and Warren's analysis \cite{Groot:1997}, $\chi_{ij}N$ was approximately linear in $\Delta a_{ij}N$ for all $N$ studied, so we used the best fit of our data $\chi_{ij} N = 0.297 \Delta a_{ij} N - 0.124$ to choose $a_{ij}$ from $\chi_{ij}N$.

\begin{figure}
    \centering
    \includegraphics{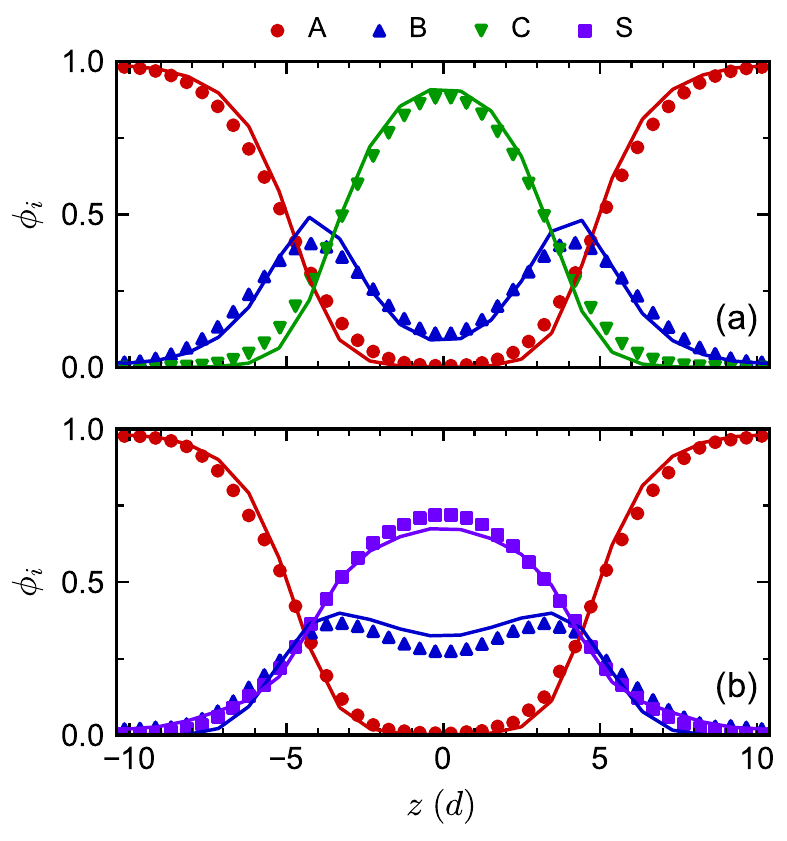}
    \caption{Volume fraction profiles from DPD (points) and SCFT (lines) in lamellar morphologies as a function of position $z$ in the direction normal to the surface (a) before and (b) after solvation when $f_{\rm A} = 0.50$ and $f_{\rm B} = 0.20$.}
    \label{fig:dpdvol}
\end{figure}

We then created lamellar morphologies of the ABC triblock with $f_{\rm A} = 0.5$ and $0 \le f_{\rm B} \le 0.2$ in an orthorhombic simulation box with square cross section (edge length $50\,d$) and using the lamellar spacing computed in the SCFT calculations (about $4.66\,R_{\rm g}$ or $20.8\,d$). We first simulated the morphologies for $5 \times 10^4\,\tau$. Then, we followed a procedure analogous to the SCFT calculations of freezing the A-block, removing the C-block, and adding solvent. We first ``froze'' all A beads and any B or C beads that were in the A-rich region of the lamella (defined as being $\ge 9\,d$ from the center of the C-rich region) by setting their velocities to zero. We then converted all unfrozen C beads to solvent (S) beads, and we removed all bonds between S beads and between S and B beads. The DPD repulsion parameter for the A and S beads was chosen as $a_{\rm AS} = 82\,k_{\rm B}T/d$ based on $\chi_{\rm AS}$ using Groot and Warren's fit \cite{Groot:1997}, while we used $a_{\rm BS} = 75\,k_{\rm B}T/d$ to give $\chi_{\rm BS} \approx 0$. We shifted the center-of-mass velocity of the unfrozen beads to zero and no longer integrated the equations of motion for the frozen beads. We then simulated the unfrozen beads for $5 \times 10^4\,\tau$, which allowed the B-block to relax against the solvent as in the SCFT calculations. The volume fraction profiles---computed from configurations sampled every $100\,\tau$ during the second half of each simulation using the same procedure as for determining $\chi_{ij}$---were in excellent agreement between DPD and SCFT for all lamellar morphologies studied, both before [Fig.~\ref{fig:dpdvol}(a)] and after [Fig.~\ref{fig:dpdvol}(b)] solvation.

After preparing the solvated membranes, we measured the local diffusivity $D(z)$ of S beads parallel to the pore surface as a function of position $z$ along the axis normal to the pore surface. Computing the solute-tracer diffusivity from the solvent-bead diffusivity is consistent with our simplifying assumption that the solute is chemically similar to the solvent; an additional bead type could be easily introduced if the solute were chemically different. We computed the parallel mean-squared displacement (MSD) of the S beads $\langle \Delta r_\parallel^2(t) | z_0 \rangle$ based on their initial $z$-position $z_0$ using spatial bins of width $1.0\,d$. We retained only beads that remained in their initial bin at time $t$ in the average.\cite{doi:10.1021/jp0375057} The local diffusivity can be extracted from the time derivative of the MSD ${\rm d}\langle \Delta r_\parallel^2 | z_0 \rangle/{\rm d}t \sim 4 D(z_0)$ once the diffusive regime is reached; however, most particles tended to diffuse out of their bins before this point. To address this, we selected a small fraction of the S beads in each bin (75 beads or about 1\%) as explicit tracers and tethered their $z$ coordinate to the center of the bin using a harmonic potential with spring constant $64\,k_{\rm B}T/d^2$. The number of tracers and strength of this potential was chosen based on simulations of the bulk solvent so that the number of tracers that remained in their initial bin increased during the relevant measurement window without significantly perturbing the system. We equilibrated the restrained system for $10^4\,\tau$, simulated for another $5 \times 10^4\,\tau$ and sampled solvent bead configurations every $0.1\,\tau$, then extracted the local diffusivity $D(z)$ from the average value of ${\rm d}\langle \Delta r_\parallel^2 | z_0 \rangle/{\rm d}t$ in the time window $500\,\tau$ to $1000\,\tau$.

To establish a point of reference for diffusivity in the membrane, we also simulated the bulk diffusivity $D_0$ of the S beads. We equilibrated the solvent in a cubic simulation box with edge length $40\,d$ for $10^3\,\tau$, then sampled configurations every $10\,\tau$ during a $5 \times 10^4\,\tau$ production simulation. We computed the three-dimensional MSD $\langle \Delta r^2(t) \rangle$ of all S beads, and we extracted the long-time diffusion coefficient from the average of its long-time derivative, ${\rm d}\langle \Delta r^2 \rangle/{\rm d}t \sim 6 D_0$, in the time window $2500\,\tau$ to $5000\,\tau$. We will report all values of the diffusivity in the membranes relative to $D_0$.

\begin{figure}
    \centering
    \includegraphics{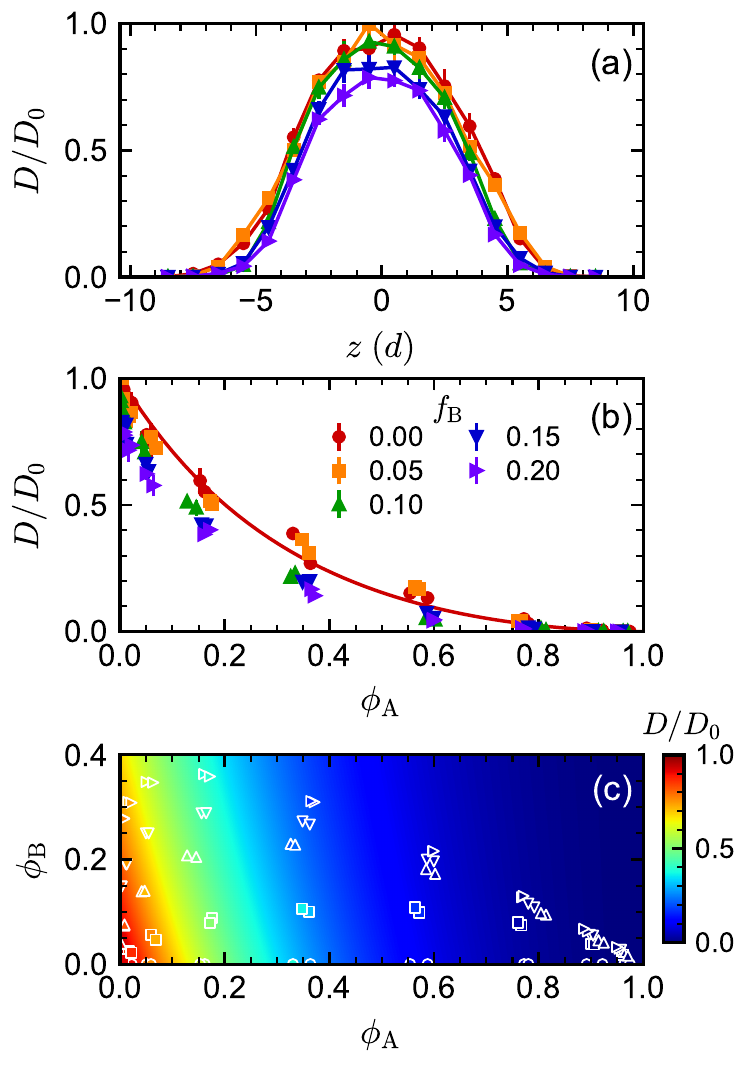}
    \caption{(a) Local diffusivity $D(z)$ in lamellar morphologies with $f_{\rm A} = 0.50$ and varied $f_{\rm B}$ from DPD simulations. (b) The same data as a function of local A-block volume fraction $\phi_{\rm A}$ for varied $f_{\rm B}$. (c) The same data as a function of $\phi_{\rm A}$ and local B-block volume fraction $\phi_{\rm B}$. In (c), the points are colored according to the measured diffusivity and the background shows the fit to Eq.~\eqref{eq:Dmodel}. The line in (b) is the fit drawn for $\phi_{\rm B} = 0$.}
    \label{fig:dpddiff}
\end{figure}

As in our previous work, the local diffusivity was highest in regions of lower polymer concentration and lowest near the pore surfaces [Fig.~\ref{fig:dpddiff}(a)], which have higher polymer concentration. In prior work, some of us showed that changes in $D(z)$ could be described by treating the B-block in the pore as a Brinkman medium with a mesh size set by the polymer correlation length \cite{Dipak2020}. We found that this picture was unable to fully describe our new simulations, which we suspected was due to the polymer model used here producing rougher pore surfaces that created additional obstructions to the solvent. This surface roughness is captured primarily in the A-block concentration and not the B-block concentration [Fig.~\ref{fig:dpddiff}(b)]. We attempted to modify the Brinkman model to also include the A-block concentration but ultimately found the fit unsatisfactory, possibly due to fundamental differences in the obstructions created by the frozen A-block and dynamic B-block. Accordingly, we posited an empirical model that treated the obstruction from the frozen A-block using a Mackie--Meares-type expression\cite{Mackie1955} and the obstruction from the dynamic B-block using our Brinkman model for an ideal polymer chain \cite{Dipak2020}:
\begin{equation}
D(\phi_{\rm A},\phi_{\rm B}) = D_0 \left(\frac{1-\phi_{\rm A}}{1+\phi_{\rm A}}\right)^m \left(1 + c \phi_{\rm B} + \frac{c^2 \phi_{\rm B}^2}{9} \right)^{-1}
\label{eq:Dmodel}
\end{equation}
We fit the parameters $m = 1.70$ and $c = 1.12$; $m$ is close to the theoretical exponent of 2 for the standard Mackie--Meares model \cite{Mackie1955}, while $c$ is a fitting parameter accounting for the hydrodynamic radius of the tracer and the scaling prefactor of the polymer correlation length \cite{Dipak2020,Rubinstein}. This empirical model for $D$ was able to fit all our lamellar measurements well [Fig.~\ref{fig:dpddiff}(c)], so we extrapolated it to the more complex membrane morphologies by assuming $D(\mathbf{r}) = D(\phi_{\rm A}(\mathbf{r}),\phi_{\rm B}(\mathbf{r}))$; this assumption neglects surface curvature effects on $D$.

\begin{figure*}
    \centering
    \includegraphics{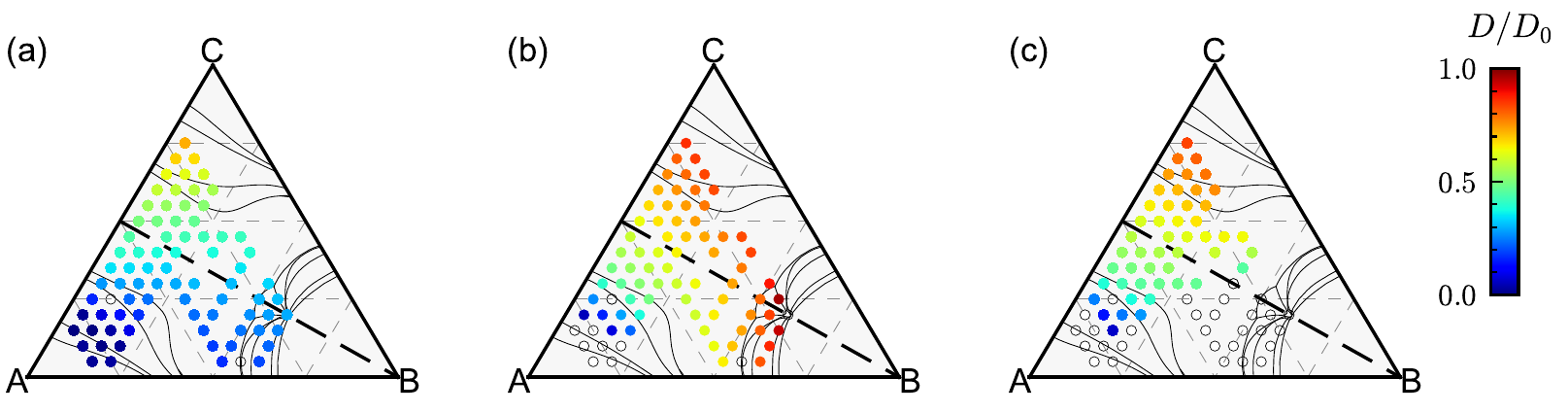}
    \caption{The average diffusivty $D$ of a solute tracer in the solvated morphologies using the (a) new kMC model, (b) RW-B+S model with pore definition $\phi_{\rm B} + \phi_{\rm S} \ge 0.5$ \cite{Howard2020}, and (c) RW-S model with pore definition $\phi_{\rm S} \ge 0.5$. The block fractions $f_i$ are those of the original triblock copolymer. Open circles indicate calculations that were attempted but could not be completed for the model because the lattice sites accessible to the tracer did not percolate for any of the 6 morphologies considered. The solid lines indicate the equilibrium phase boundaries \cite{Tyler2007}.}
    \label{fig:ternary}
\end{figure*}

\subsection{Kinetic Monte Carlo simulation details}
To simulate diffusion using the kMC model, we discretized the membrane morphology onto the same lattice as was used in the SCFT calculations. (A smaller lattice spacing was tested for selected configurations and found not to significantly impact the results.) We computed $\psi$ using $\phi_{\rm S}$, and $D$ using $\phi_{\rm A}$ and $\phi_{\rm B}$ at each lattice site. To ensure there was no diffusion into regions of large $\phi_{\rm A}$, which should be negligible in both experiments and the DPD simulations, we set the external field to $\beta\psi = \infty$ when $\phi_{\rm S} < 0.02$ to effectively disallow moves to these sites (make $k$ zero). We then computed the kMC hopping rates according to Eq.~\eqref{eq:kmcrate} using a second-order central finite difference scheme to estimate the required gradients. Care was taken to use an appropriate forward or backward difference when $\beta\psi = \infty$ at a neighboring site.

To carry out the moves, we employed a rejection-free scheme that randomly selected a move and randomly advanced the time coordinate for each tracer according to the rates at a given lattice site.\cite{GILLESPIE1976403, Gillespie1977, Chatterjee2007} For a tracer at lattice site $i$, we randomly chose to move to an adjacent lattice site $i_\alpha^\pm$ with the weight of selecting each site being proportional to $k(i_\alpha^\pm|i)$. We unconditionally accepted this move and advanced the time coordinate for each tracer by a random amount that was exponentially distributed with mean $[\sum k(i_\alpha^\pm|i)]^{-1}$, where the sum is over all adjacent sites of $i$. We found that this rejection-free scheme was computationally advantageous compared to a scheme that advanced the time by a fixed amount but allowed move rejection, i.e., the tracer could remain at site $i$ \cite{Chatterjee2007}. The rejection scheme was roughly twice as fast as the rejection-free scheme per move because it required half as many random numbers, but the rejection-free scheme was ultimately faster overall when the move rejection rate exceeded 50\%. This rejection rate was quite common in the nonequilibrium morphologies, where the average tracer diffusivity was significantly less than the nominal bulk diffusivity $D_0$.

Using this scheme, we simulated an ensemble of $5 \times 10^{4}$ tracers that we randomly initialized onto all lattice sites for which $\beta\psi \ne \infty$. We simulated for $10^4\,\tau$ to allow the tracers to relax to their equilibrium distribution, then sampled the tracer coordinates every $100\,\tau$ during a $10^5\,\tau$ simulation. We computed $\langle \Delta r^2 \rangle$ for all tracers and extracted $D$ from ${\rm d}\langle \Delta r^2 \rangle/{\rm d}t$ in the time window $2 \times 10^4\,\tau$ to $5 \times 10^4\,\tau$.

\section{Results and discussion}
\subsection{Characterization of models for diffusion}
Having developed our multiscale model, we proceeded to analyze the model's predictions and compare them to alternative approaches. In particular, some of us previously used a simple random-walk (RW) model for solute diffusion in triblock copolymer membranes \cite{Howard2020}. Unlike the approach described in Section~\ref{sec:method:kmc}, the RW model adopted a binary definition of the pores based on the local volume fraction of the A-block matrix $\phi_{\rm A}$: sites with $\phi_{\rm A} < 0.5$ (or $\phi_{\rm B} + \phi_{\rm S} \ge 0.5$) were defined as the pores, and all other sites were defined as matrix inaccessible to the solute. Within the pores, variations in the distribution of the solute and the local diffusivity were both neglected. This RW model is effectively a special case of the kMC approach assuming $\beta\psi = 0$ and $D/D_0 = 1$ inside the pores, and $\beta\psi = \infty$ outside the pores. The RW model is simpler than our new approach because it does not require the additional SCFT calculation to solvate the B-block inside the pore; however, the RW model is potentially less accurate because it neglects pore-level effects on the solute distribution and transport \cite{Dipak2020}. Hence, we compared our diffusion measurements using the new approach that incorporates these effects (Sec.~\ref{sec:method}) to the RW model. We will refer to our new approach as the ``kMC model'' and the approach of Ref.~\citenum{Howard2020} as the ``RW-B+S model''; the latter emphasizes that the pore definition in that model includes both the B block and the solvent.

We first considered the average solute diffusivity $D$ as a function of the polymer block fractions $f_i$ (Fig.~\ref{fig:ternary}) for the nonequilibrium morphologies of Ref.~\citenum{Howard2020} (Sec.~\ref{sec:method:scft}). We excluded all morphologies with $f_{\rm A} \le 0.2$ because they tended to produce matrices (using the RW-B+S definition of pore and matrix sites) that were not well connected and so were not mechanically viable for membrane applications. This was not surprising given the limited content of matrix-forming A block and that the equilibrium self-assembled structures for these polymers, such as sphere or disordered phases, also are not connected. We further excluded any morphologies where the lattice sites accessible to the tracer (within each model) were not percolated through the periodic boundaries in at least one dimension, as these morphologies would produce $D/D_0 = 0$ at long times. Figure \ref{fig:ternary}(a) shows the diffusivity predicted by the kMC model, while Fig.~\ref{fig:ternary}(b) shows the same for the RW-B+S model. As a verification of our approach, we note that we achieved nearly quantitative agreement between Fig.~\ref{fig:ternary}(b) and the data of Ref.~\citenum{Howard2020} (largest difference in $D/D_0$ of 0.033). This agreement helps confirm that our method for constraining the matrix-forming A block is effective because simulations of the RW-B+S model using the solvated morphologies should produce identical results to simulations using the original melt morphologies of Ref.~\citenum{Howard2020} if $\phi_{\rm A}(\mathbf{r})$ is successfully frozen.

\begin{figure}
    \centering
    \includegraphics{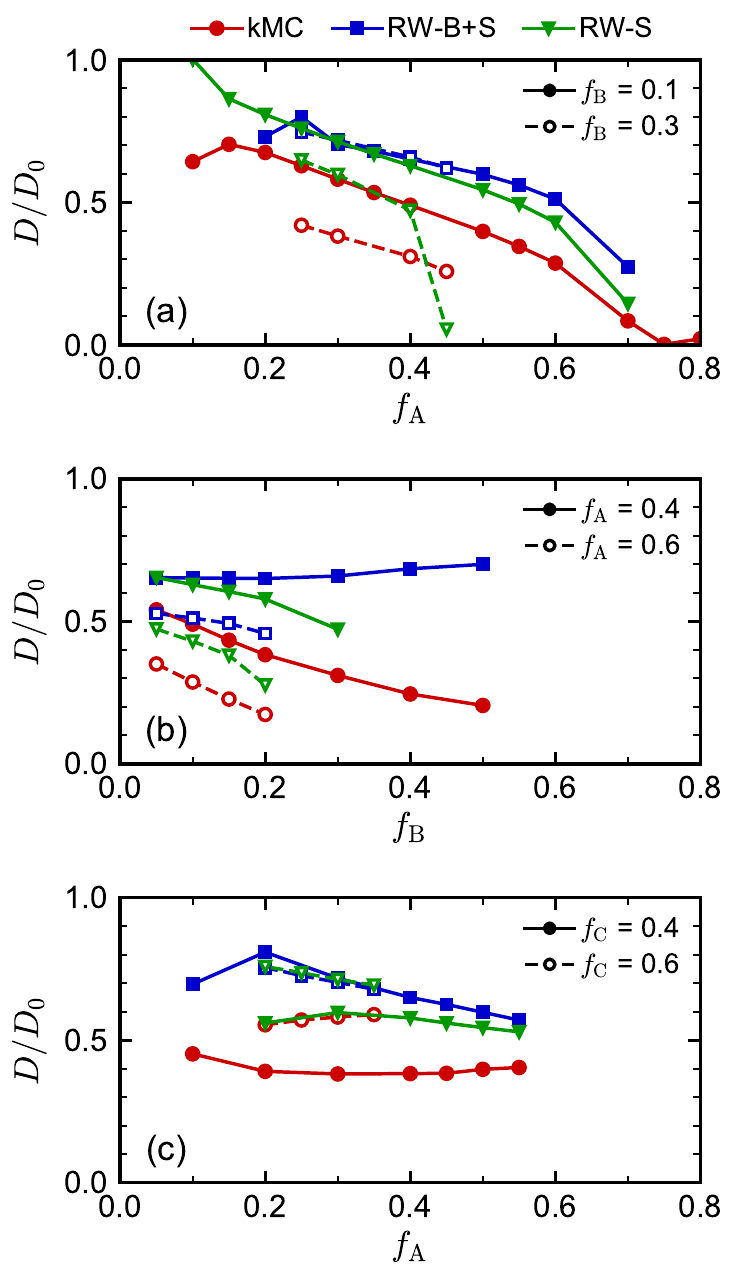}
    \caption{The average diffusivity $D$ of a solute tracer from Fig.~\ref{fig:ternary} projected along lines of constant (a) $f_{\rm B}$, (b) $f_{\rm A}$, and (c) $f_{\rm C}$ using the new kMC model, RW-B+S model, and RW-S model. The block fractions $f_i$ refer to the original triblock copolymer.}
    \label{fig:constant}
\end{figure}

In general, $D$ increased in both models as $f_{\rm A}$ decreased due to the increased space available for the tracer to diffuse. However, the diffusivity simulated using the kMC model was smaller than that simulated using the RW-B+S model, particularly for smaller $f_{\rm A}$, because the local diffusivity $D(\mathbf{r})$ input to the kMC model was typically less than $D_0$ (Fig.~\ref{fig:dpddiff}). To more easily visualize these trends, we projected $D$ as a function of $f_{\rm A}$ along lines of constant $f_{\rm B}$ [Fig.~\ref{fig:constant}(a)], which clearly showed a decrease in $D$ with respect to $f_{\rm A}$ for both models but consistently smaller values of $D$ in the kMC model than in the RW-B+S model.

An even more striking difference between the two models was the dependence of $D$ on $f_{\rm B}$. Based on prior DPD simulations \cite{Dipak2020}, we expected the diffusivity to decrease with increasing $f_{\rm B}$ because the B block obstructs solute motion in the pores. The kMC model produced diffusivities consistent with this expectation [Fig.~\ref{fig:ternary}(a)], but the RW-B+S model did not [Fig.~\ref{fig:ternary}(b)]. For example, in Fig.~\ref{fig:constant}(a), the diffusivities obtained using the kMC model were smaller when $f_{\rm B} = 0.3$ than when $f_{\rm B} = 0.1$, but the diffusivities obtained using the RW-B+S model were essentially the same. To more clearly highlight this behavior, we projected $D$ as a function of $f_{\rm B}$ along lines of constant $f_{\rm A}$ [Fig.~\ref{fig:constant}(b)]. For the kMC model, there was a clear monotonic decrease in $D$ as a function of $f_B$; however, for the RW-B+S model, there was very little variation in $D$ with $f_{\rm B}$. In fact, we saw a slight increase in $D$ at the largest values of $f_{\rm B}$ when $f_{\rm A}=0.4$. This difference is a consequence of the kMC model incorporating a spatially varying local diffusivity $D(\mathbf{r})$ that depends on the local polymer concentration in the pores rather than the constant diffusivity assumed in the RW-B+S model; we emphasize that we consider the former to be more realistic.

Furthermore, projecting the diffusivity predicted by the kMC model along lines of constant $f_{\rm C}$ revealed that $D$ was roughly constant with respect to changes in $f_{\rm A}$ [Fig.~\ref{fig:constant}(c)]. In order to maintain fixed $f_{\rm C} = 1-f_{\rm A}-f_{\rm B}$, $f_{\rm B}$ must decrease as $f_{\rm A}$ increases. The local diffusivity in the pores depends on the composition, and the exchange of A for B can lead to a direct competition between the effects of these blocks on the diffusivity. For example, increasing $f_{\rm A}$ tends to decrease $D$ by decreasing the matrix volume that is accessible to the solute; however, the necessary accompanying decrease in $f_{\rm B}$ at constant $f_{\rm C}$ leads to less B in the pores and tends to increase $D$. In contrast, in the RW-B+S model, the pore size decreases as $f_{\rm A}$ increases at constant $f_{\rm C}$ but there is no accompanying increase in the diffusivity inside the pores (due to the decrease in $f_{\rm B}$), resulting in more obstructed motion and smaller $D$. This compensating behavior may be sensitive to the solute and polymer chemistry ($\chi_{ij}$) as well as the model for the solute diffusivity.

We also noted that there were some exceptions to this behavior. When $f_{\text{C}} \le 0.2$, we no longer observed constant diffusivity along lines of constant $f_{\text{C}}$ in the kMC model, primarily when $f_{\rm A}$ was large [Fig.~\ref{fig:ternary}(a)]. Increasing $f_{\rm A}$ to large values led to fragmentation of the pores due to the formation of larger A domains, reducing the accessible pathways for diffusion. In this case, exchanging $f_{\rm A}$ for $f_{\rm B}$ at constant $f_{\rm C}$ led to an overall increase in diffusivity. In membrane applications, exchanging the fractions of A and B may also have practical implications on separation performance that are not directly captured by $D$. For example, if the A block is hydrophobic and the B block is hydrophilic, increasing the B block content may increase water uptake. Moreover, increasing $f_{\rm A}$ may lead to an increase of dead-ends in nonequilibrium morphologies that hamper directed transport across the membrane \cite{Schneider2019}.

After comparing our new approach to our prior work, we asked whether we could combine features of both to construct a RW model that captured similar trends in $D$ as the kMC model but maintained the simplicity of the RW approach. The B-block has two important effects in the kMC model: it excludes the solute from parts of the pore through $\psi$ and it obstructs diffusion in the pore through $D(\mathbf{r})$. The RW model can approximately treat the first effect by redefining the pores. We proposed an alternative model definition where sites with $\phi_{\text{A}} + \phi_{\text{B}} < 0.5$ (or $\phi_{\rm S} \ge 0.5$) were considered to be the pores, and all other sites were the matrix. This should be considered a crude but convenient approximation because even dense regions of B were typically partially permeable to the solute in all our models. We will refer to this model as the ``RW-S'' model to emphasize that the pore definition includes essentially only the solvent.

Similar to our results with the kMC and RW-B+S models, the RW-S model captured the monotonic decrease in $D$ as a function of $f_{\rm A}$ [Figs.~\ref{fig:ternary}(c) and \ref{fig:constant}(a)]. Notably, the RW-S model also gave values of $D$ that decreased with increasing $f_{\rm B}$ [Fig.~\ref{fig:constant}(b)], indicating that effects of the B-block can be at least partially approximated as an additional obstruction. However, $D$ was not constant with respect to $f_{\rm A}$ at constant $f_{\rm C}$ [Fig.~\ref{fig:constant}(c)] in the RW-S model. This finding is consistent with our hypothesis that $D$ is nearly constant in the kMC model along lines of constant $f_{\rm C}$ because of a tradeoff between the A-block and B-block in the local diffusivity; the RW-S model has constant local diffusivity, so this tradeoff cannot be captured. We also note a practical drawback of the RW-S model: because the pore definition is more restrictive than that in the RW-B+S model, we were unable to run simulations for many of the polymers with large $f_{\rm A}$ or $f_{\rm B}$. Our kMC model, on the other hand, did not suffer from this because an artificial (binary) classification of the lattice sites as pore or matrix was not required. In this respect, we view our new multiscale approach based on the kMC model as being more convenient, systematic, and faithful to physical expectations than either of the RW models.

\subsection{Correlating diffusion with structure}
In Ref.~\citenum{Howard2020}, we showed that the diffusivity of the RW-B+S model correlated strongly with two structural descriptors of the pores: namely, their volume $v$ and integrated mean curvature $h$, normalized by the total membrane volume. These two descriptors, along with the normalized surface area $s$ and integrated Gaussian curvature $g$, comprise the four Minkowski functionals from integral geometry and image analysis \cite{MICHIELSEN2001, Armstrong2019}. Given the qualitative differences we observed between the kMC model and the RW-B+S model, we asked whether the diffusivity of the kMC model correlated with any of these functionals.

The Minkowski functionals can be computed from black-and-white (binary) images using a voxel-counting algorithm \cite{MICHIELSEN2001}. The RW-B+S model has a binary pore definition suitable for this algorithm, but the kMC model does not because the solute can access most lattice sites but diffuses at different rates through them. We accordingly must adopt an additional structural definition for the pores in the kMC model, and two natural choices are that of either the RW-B+S model or the RW-S model. Given that the RW-S model did not give percolated pores at many state points [Fig.~\ref{fig:ternary}(c)], we chose to use the RW-B+S model to define the pores. Hence, the Minkowski functionals we computed were the same for a given morphology for both the kMC and RW-B+S models, but the corresponding diffusivities were different.

\begin{figure*}
    \centering
    \includegraphics{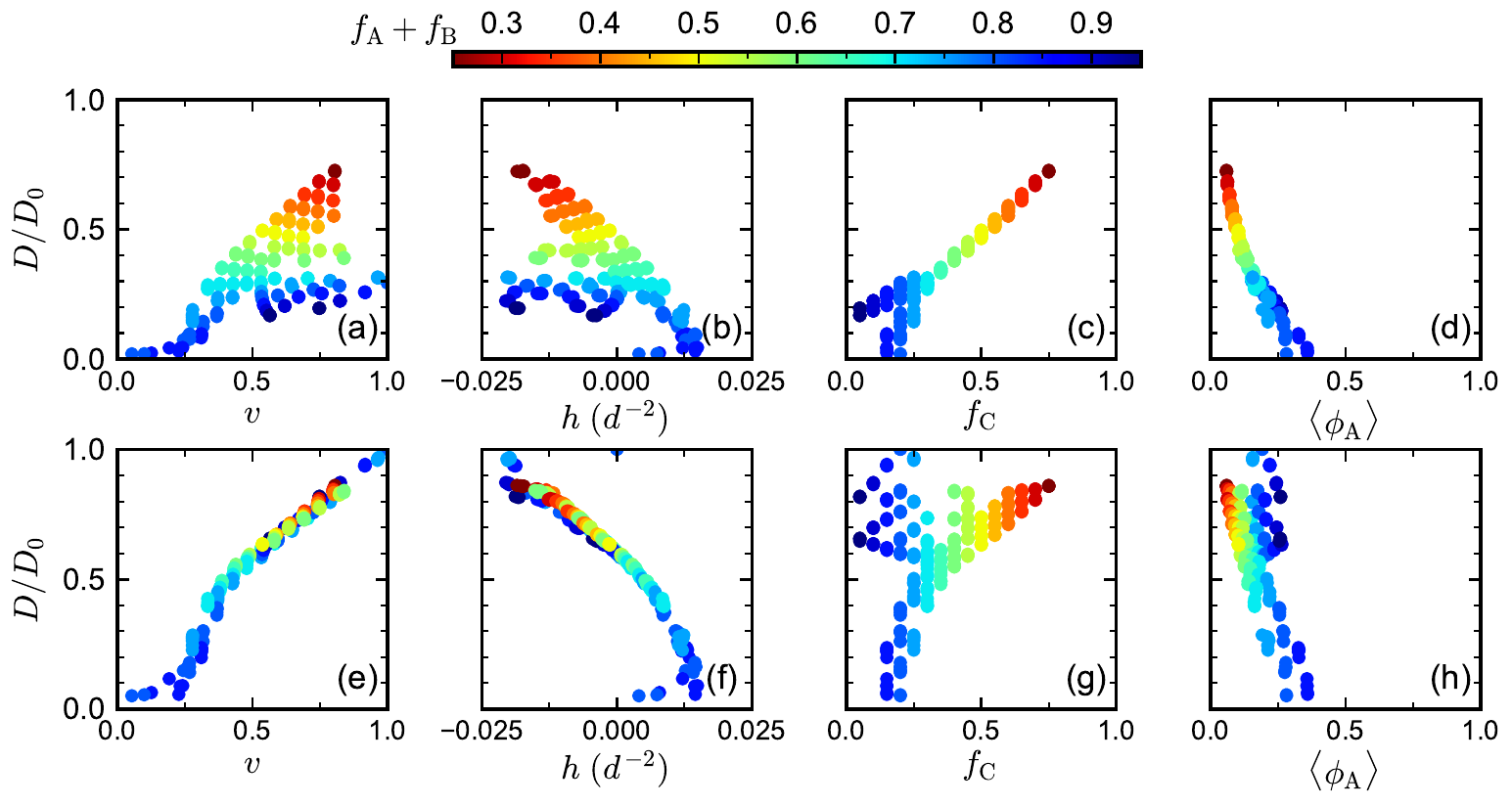}
    \caption{Correlation of the diffusivity $D$ of a solute tracer calculated by the (a)--(d) kMC model and (e)--(f) RW-B+S model with various selected structural descriptors: the pore volume $v$ and integrated mean curvature $h$ (normalized by the total volume of the membrane), the C-block fraction of the triblock copolymer $f_{\rm C}$, and the average A-block volume fraction experienced by the solute $\langle \phi_{\rm A} \rangle$. Data are shown for all morphologies that percolated in at least one dimension.}
    \label{fig:featcorr}
\end{figure*}

We plotted $D$ against the four Minkowski functionals for both models (Fig.~\ref{fig:featcorr}). In sharp contrast to the RW-B+S model [Fig.~\ref{fig:featcorr}(e)--(f)] \cite{Howard2020}, $D$ obtained using the kMC model did not correlate strongly with the volume $v$ or integrated mean curvature $h$ [Fig.~\ref{fig:featcorr}(a)--(b)]. The relationship between $D$ and $v$ was no longer approximately one-to-one: there were many values of $v$ that gave similar values of $D$. We rationalized this as being due to the B-block coating the pores. Within the RW-B+S model, the pore volume does not change as the length of the pore coating increases and neither does the diffusivity; however, the diffusivity computed using the kMC model does change [Fig.~\ref{fig:constant}(b)]. Similar differences were observed for correlation with $h$ between the two models. Neither the kMC model nor the RW-B+S model showed correlation between $D$ and the surface area $s$ or integrated Gaussian curvature $g$ (not shown).

Motivated by this lack of correlation between $D$ and the Minkowski functionals for the kMC model, we considered additional structural descriptors of the polymer or membrane that might capture variations in $D$. The simplest descriptors we added were the polymer block fractions $f_i$, which we chose because of the trends observed in Figs.~\ref{fig:ternary} and \ref{fig:constant}. Only two of these block fractions are independent parameters, so we focus our discussion on $f_{\rm B}$ and $f_{\rm C}$. The diffusivity was uncorrelated with $f_{\rm B}$ for both models (not shown), which we attribute to the inability of $f_{\rm B}$ to capture major structural changes in the morphology as $f_{\rm A}$ or $f_{\rm C}$ is varied for the triblock copolymer we studied. Interestingly, $D$ correlated well with $f_{\rm C}$ in the kMC model for sufficiently large $f_{\rm C}$ [Fig.~\ref{fig:featcorr}(c)] although it was essentially uncorrelated with $f_{\rm C}$ in the RW-B+S model [Fig.~\ref{fig:featcorr}(g)]. We noted, however, that the correlation broke down at smaller values of $f_{\rm C}$ where the pores tended to be less percolated. The block fractions are convenient descriptors based on solely the triblock copolymer architecture, but they do not capture the actual self-assembled membrane morphology or the environment experienced by the solute inside the pores.

Given that the local diffusivity is a function of the local composition [Eq.~\eqref{eq:Dmodel}], we posited two additional descriptors for the solute environment to help improve on the block-fraction descriptors. In particular, we computed the average polymer volume fractions experienced by the solute
\begin{equation}
    \langle \phi_{i} \rangle = \int {\rm d}\mathbf{r} \ \phi_{i}(\mathbf{r}) \rho_\infty(\mathbf{r}),
\end{equation}
where $\rho_\infty \sim \phi_{\rm S}$ is the steady state probability distribution of the solute, encoded in the kMC model by $\psi$. We considered both $\langle \phi_{\rm A} \rangle$ and $\langle \phi_{\rm B} \rangle$. For the RW-B+S model, we found that $D$ did not correlate strongly with either $\langle\phi_{\rm A}\rangle$ [Fig.~\ref{fig:featcorr}(h)] or $\langle\phi_{\rm B}\rangle$ (not shown); we somewhat expected this result because the local diffusivity in the RW-B+S model did not depend on the local polymer concentration, so the long-time diffusivity depends primarily on the pore morphology and this dependence is better captured by other descriptors. In the kMC model, $D$ had a strong correlation with $\langle\phi_{\rm A}\rangle$ [Fig.~\ref{fig:featcorr}(d)] but not $\langle\phi_{\rm B}\rangle$ (not shown).

\begin{figure}
    \centering
    \includegraphics{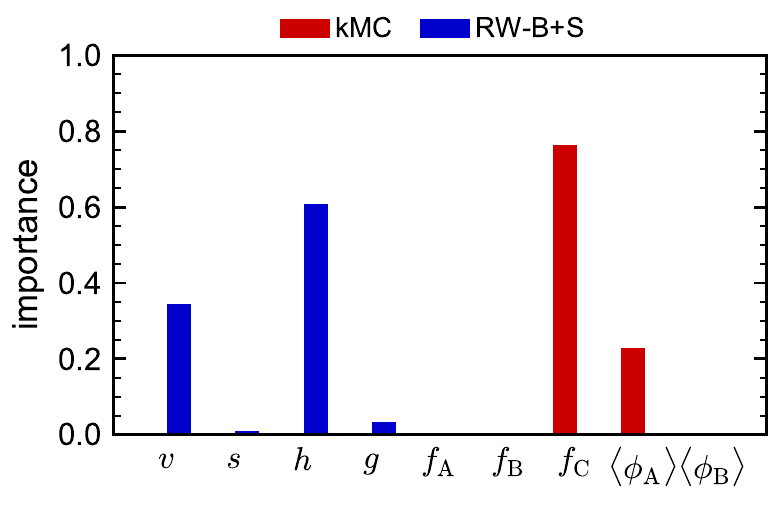}
    \caption{Relative importance of each feature in a random forest regression for diffusivity calculated by the kMC model and RW-B+S model.}
    \label{fig:importance}
\end{figure}

In order to quantify which structural descriptors correlate most strongly with the solute diffusivity for each model, we used the same approach as Ref.~\citenum{Howard2020} and regressed $D$ as a function of the four Minkowski functionals, the three block fractions, and $\langle \phi_{\rm A}\rangle$ and $\langle \phi_{\rm B} \rangle$ using a random-forest model \cite{Breiman2001, Pedregosa2011}. To determine the importance of each descriptor, we randomized the values of each descriptor and measured the resulting mean decrease in accuracy of the model. The importance of each descriptor for each model (Fig.~\ref{fig:importance}) was determined from the relative magnitude of its mean decrease in accuracy, such that the sum of importances was 1. The results are largely consistent with our visual observations: $v$ and $h$ were the most important descriptors for the RW-B+S model, while $f_{\rm C}$ and $\langle \phi_{\text{A}} \rangle$ were the dominant features for the kMC model. In the kMC model, we found that $f_{\text{C}}$ was roughly three times more important than $\langle \phi_{\text{A}} \rangle$, which we interpreted as being a result of $f_{\rm C}$ capturing most of the variation in $D$ except for polymers having sufficiently small $f_{\rm C}$. It is interesting that the importance of these descriptors changes significantly between the two transport models.

These results, obtained using our new multiscale model that accounts for pore-level effects on transport, seem to reveal vastly different trends in diffusivity compared to a simpler model that neglected these effects \cite{Howard2020}, providing new insights into the features that may be important in designing triblock copolymer membranes with nonequilibrium morphologies. We find that the diffusivity of solvent through the pores is mainly impacted by the length of the sacrificial block and the average contact the solvent has with the porous walls. The latter quantity is not readily accessible in experiments, however, it fortunately appears to be most important only when the degree of percolation of the pores is limited. Despite failing to find a direct correlation relationship between diffusivity and structural descriptor based on the Minkowski functionals, we note that $f_{\rm C}$ does ultimately play a role in forming the structure in a way not easily captured by these functionals. Furthermore, we emphasize that our set of Flory--Huggins interaction parameters only roughly approximates one triblock copolymer--solvent system, and different interaction parameters may yield alternative or additional correlations.

\section{Conclusions}
We have developed a multiscale model to study solute diffusion through a porous triblock copolymer membrane where the pores have a brush-like coating that interacts with the solute. The method uses SCFT to simulate self-assembly and solvation of the membrane, with the latter implemented using a novel field theory to constrain a target density profile. Then, on-lattice kMC was used to simulate diffusion under the influence of a spatially varying external field and local diffusivity describing the solute interactions with the membrane. We applied our model to simulate solute diffusion through nonequilibrium morphologies of a model ABC triblock copolymer, revealing vastly different trends in diffusivity compared to a simpler geometric model that neglected solute--membrane interaction effects \cite{Howard2020}. In particular, we found that the diffusivity of solvent through the pores was mainly impacted by the length of the sacrificial polymer block used to create the pores as well as the average contact the solute had with the porous walls. The latter quantity is not readily accessible in experiments; however, it appeared most important when the degree of percolation of the pores was limited. Our results imply that transport properties of nonequilibrium triblock copolymer membranes might be controlled by straightforward tailoring of the polymer block lengths, thus alleviating the need to target a particular structure or to attempt to minimize defects. 

The model system studied here consists of a nonfrustrated ABC triblock copolymer, a water-like solvent that has strong hydrophobic and hydrophilic interactions with the membrane-matrix A block and pore-coating B block, respectively, and a model solute. The Flory--Huggins interaction parameters $\chi_{ij}$ that we have employed hence correspond to only one possible realization of this system. Our multiscale workflow can be easily applied to study specific polymer, solvent, and solute chemistries with appropriate choice of $\chi_{ij}$ for the SCFT simulations and a model for the local diffusivity, which might be obtained from experimental data or molecular simulations. We also expect that the model could be straightforwardly extended to incorporate other interactions between the solute and membrane, such as electrostatics.

\begin{acknowledgements}
This work was supported as part of the Center for Materials for Water and Energy Systems (M-WET), an Energy Frontier Research Center funded by the U.S. Department of Energy, Office of Science, Basic Energy Sciences under Award \#DE-SC0019272. VG and TMT acknowledge financial support from the Welch Foundation (Grant Nos. F-1599 and F-1696). SK acknowledges support from the National Science Foundation through the Center for Dynamics and Control of Materials: an NSF Materials Research Science and Engineering Center (NSF MRSEC) under Cooperative Agreement DMR-1720595. We acknowledge the Texas Advanced Computing Center (TACC) at The University of Texas at Austin for providing HPC resources. Use was made of computational facilities purchased with funds from the National Science Foundation (CNS-1725797) and administered by the Center for Scientific Computing (CSC). The CSC is supported by the California NanoSystems Institute and the Materials Research Science and Engineering Center (MRSEC; NSF DMR-1720256) at UC Santa Barbara.
\end{acknowledgements}

\section*{Data Availability}
The data that support the findings of this study are available from the authors upon reasonable request.

\bibliography{bibliography}

\end{document}